\documentclass[useAMS,usenatbib]{mn2e}

\usepackage{graphicx}
\usepackage{subfigure}
 \usepackage{graphicx,amsmath,amssymb}
\usepackage{color} 
\usepackage{hyperref}
\usepackage{color}
\usepackage{amsmath}
\def\lsim{\lower.5ex\hbox{$\; \buildrel < \over \sim \;$}}
\def\gsim{\lower.5ex\hbox{$\; \buildrel > \over \sim \;$}}
\def\simeq{\lower.3ex\hbox{$\; \buildrel \sim \over - \;$}}
\def\ch{\lower-0.55ex\hbox{--}\kern-0.55em{\lower0.15ex\hbox{$h$}}}
\def\lh{\lower-0.55ex\hbox{--}\kern-0.55em{\lower0.15ex\hbox{$\lambda$}}}

\def\lsim{\lower.5ex\hbox{$\; \buildrel < \over \sim \;$}}
\def\gsim{\lower.5ex\hbox{$\; \buildrel > \over \sim \;$}}
\def \simeq{\lower.3ex\hbox{$\; \buildrel \sim \over - \;$}}

\title[Shocks in magnetized accretion flows]
{Standing shocks in magnetized advection accretion flows onto a rotating black hole}
\author[Santabrata Das, Biplob Sarkar]
{Santabrata Das\thanks{E-mail: sbdas@iitg.ernet.in; biplob@iitg.ernet.in},
Biplob Sarkar\footnotemark[1] \\
Indian Institute of Technology Guwahati, Guwahati, 781039, India.
}
\begin{document}

\date{Accepted . Received ; in original form }

\pagerange{\pageref{firstpage}--\pageref{lastpage}} \pubyear{}

\maketitle

\label{firstpage}

\begin{abstract}

We present the global structure of magnetized advective accretion flow around the
rotating black holes in presence of dissipation. By considering accretion flow to
be threaded by toroidal magnetic fields and by assuming synchrotron radiative
mechanism to be the dominant cooling process, we obtain global transonic accretion 
solutions in terms of dissipation parameters, such as viscosity ($\alpha_B$), 
accretion rate (${\dot m}$) and plasma-$\beta$, respectively. In the rotating magnetized 
accretion flow, centrifugal barrier is developed in the nearby region of the black hole
that triggers the discontinuous shock transition in the flow variables. Evidently,
the shock properties and the dynamics of the post-shock flow (hereafter post-shock
corona (PSC)) are being governed by the flow parameters. We study the role of
dissipation parameters in the formation of standing shock wave and find that 
global shocked accretion solutions exist both in gas pressure dominated 
flows and in magnetic pressure dominated flows. In addition, we observe that standing shock 
continues to form around the rapidly rotating black holes as well. We identify the range of
dissipation parameters that permits shocked accretion solutions and find that standing 
shocks continue to form even in presence of high dissipation limit, although the likelihood
of shock formation diminishes with the increase of dissipation. Further, we compute the critical 
accretion rate (${\dot m}^{\rm cri}$) that admits shock and observe that standing shock 
exists in a magnetically dominated accretion flow when the accretion rate lies in
general in the sub-Eddington domain. At the end, we calculate the maximum
dissipated energy that may be escaped from the PSC and indicate its possible implication
in the astrophysical context.

\end{abstract}

\begin{keywords}
accretion, accretion discs, magnetic field, black hole physics, magnetohydrodynamics (MHD), shock waves
\end{keywords}

\section{Introduction}

Magnetic fields are in general considered to be indispensable in the
astrophysical environment and therefore, their presence in the 
accretion disc is by all means inevitable \citep{Balbus-Hawley98}. In a magnetized accretion disc, 
magnetic fields play an important role in guiding the infalling matter around black holes. 
Meanwhile, \citet{Blandford-Payne82} revealed  that when a Keplerian disc is threaded
by large scale magnetic fields, angular momentum can be removed through the torque
exerted by the magnetic fields. 
Similarly, the large scale poloidal magnetic fields anchored in the surrounding accretion disc
are indeed capable of transferring energy and angular momentum and also instigate the generation
of powerful magnetic jets \citep{Blandford-Znajek77,Komissarov-McKinney07}. Further, 
\citet{Balbus-Hawley91,Balbus-Hawley98} showed that the accretion disc becomes unstable
in presence of differential rotation when the accreting plasma is threaded by weak vertical
magnetic fields. 
This instability causes the turbulence leading to the efficient angular momentum transport
as well as energy dissipation that enables the accretion possible.

In the modeling of the standard advection-dominated accretion flows around black holes, 
\citet{Narayan-Yi95} considered the magnetic fields which are stochastic in nature. 
However, since the flow experiences differential rotation while accreting onto a black hole, 
the magnetic fields present in the disc are expected to be structured in reality and the
large scale fields seem to be dominated by its toroidal component. This consideration
in general holds irrespective to the initial configuration of the fields ($i.e.$, toroidal
or poloidal). Furthermore, the existence of toroidal magnetic field has been 
observationally confirmed in the exterior regions of the discs of young stellar objects
\citep{Aitken-etal93,Wright-etal93} as well as in the Galactic center \citep{Chuss-etal03,
Novak-etal03}. Meanwhile, significant efforts were given to examine the accretion disc
properties around black holes including toroidal magnetic fields
\citep{Akizuki-Fukue06,Khesali-Faghei08,Khesali-Faghei09,mosall14, mosall16,
Oda-etal07,Oda-etal10,Oda-etal12,Samadi-etal14,sarkar2015,sarkar2016,
sarkar2018,sarkar2017b}. Following the above cognizance, in the present work, 
we consider the accretion flow to be threaded by toroidal magnetic field lines
as well.

Further, while developing the present formalism, we consider rotating
matter that experiences centrifugal repulsion as it accretes towards the 
black hole and due to this, infalling matter is being piled up in the vicinity
of the black hole. In reality, such
accumulation of matter can not be continued indefinitely and ultimately, at its 
limit, the centrifugal barrier triggers the discontinuous transition of the flow
variables which is commonly called as shock transition. It may be noted that
the global accretion solutions including shock waves are potentially favored as
it owns large amount of entropy \citep{Becker-Kazanas01}. In the
theoretical front, the shock induced global accretion solution around black 
hole and its implications are extensively studied by the numerous groups of
workers \citep{Fukue87,Chakrabarti89,Chakrabarti96b,Lu-etal99,Gu-Lu01,
Das-etal01b,Gu-Lu04,Fukumura-Tsuruta04,Chakrabarti-Das04,Mondal-Chakrabarti06,Das07,
Becker-etal2008,Das-etal09,Das-etal10,sarkar2015,Aktar-etal15,sarkar2016,
Aktar-etal17,sarkar2017b,sarkar2018}. In addition, the existence of shock in
accretion flow is also examined numerically considering hydrodynamics
\citep{ChakrabartiMolteni1993,Moltenietal1994,Ryuetal1997,Okuda14,
Okuda-Das15,Sukova-Janiuk15,Sukova-17} as well as magnetohydrodynamic (MHD)
environment \citep{Nishikawa-etal05,Takahashi-etal06,Fukumura-etal07,Fukumura-etal16}.

Motivated with the above studies, in this work, we examine the magnetically
supported accretion flow around rotating black hole that possesses standing shock.
While doing this, we assume that the characteristics of the magnetic pressure
is synoptic to the gas pressure and their combined effects therefore supports the
vertical structure of the infalling matter against the gravitational pull. Moreover,
recalling the success of 
the seminal $\alpha$-viscosity prescription \citep{Shakura-Sunyaev73}, we consider the
Maxwell stress to be proportional to the total pressure \citep{Machida-etal06}
that evidently demonstrates 
that the outward transport of angular momentum
would certainly be enhanced as the magnetic activity inside the disc is increased.
Furthermore, we consider the heating of the flow to be regulated by the
magnetic energy dissipation mechanism while the inflowing matter is being cooled via 
synchrotron emission process
\citep{Chattopadhyay-Chakrabarti00,Das07,sarkar2018}.
In addition, for simplicity, we adopt a pseudo potential introduced by \citet{Chakrabarti-Mondal06}
that successfully mimics the space-time geometry around the rotating black hole
having spin $a_k \lesssim 0.8$. Considering all these, we self-consistently
solve all the governing equations that describe the magnetized accretion flow around
rotating black hole and obtain the global accretion solutions including shock
waves. We study the properties of standing shock waves in terms of flow parameters
and observe that shock formation takes place for an ample range of parameters both around
weakly rotating ($a_k \rightarrow 0$) as well as rapidly rotating black holes
($a_k \sim 0.8$). We also calculate the critical accretion rate (${\dot m}^{\rm cri}$)
for standing shocks in magnetized accretion flow. 
It may be noted that ${\dot m}^{\rm cri}$ does not bear
any universal value, rather it is largely dependent on the inflow parameters. We
continue our study considering the fact that standing accretion shocks are
dissipative by nature 
and calculate the maximum energy that can be extracted from the PSC. In reality, this
available energy could be utilized in powering the jets
\citep[reference therein]{sarkar2016} as they seem to originate
from PSC regions \citep[reference therein]{Aktar-etal17}.

We organize the paper as follows. In \S 2, we write the model equations
and carry out the analysis of transonic conditions. 
In \S 3, we display our results where shocked 
accretion solutions for magnetized flow and its properties are discussed.
Moreover, we determine the critical inflow parameters for standing 
shock as well. We further study the characteristics of dissipative standing shock.
Finally, in \S 4, concluding remarks are presented.

\section{Accretion flow model}

To take into consideration of the magnetic fields structure in an accretion disc, we
rely on the numerical simulation results of global and local
MHD accretion flow around black hole. These simulations have revealed that magnetic
fields inside the accretion disc are turbulent and primarily dominated by the azimuthal
component \citep{Hirose-etal06,Machida-etal06,Johansen-Levin08}. Following
the findings of these simulations, we separate the magnetic fields into mean fields,
denoted by ${\bf{B}} = (0,<B_{\phi}>,0)$, and the fluctuating fields, indicated as
$\delta {\bf{B}}= (\delta {B}_{r}, \delta {B}_{\phi},\delta {B}_{z})$. Here, we express the
azimuthal average by `$<>$' and upon azimuthal averaging, the fluctuating components
of the magnetic fields eventually disappear ($<\delta {\bf{B}}> = 0$). Moreover, 
the radial and vertical components of the magnetic field are assumed to be negligible
when compared with the azimuthal component,
$\mid <B_{\phi}> + \delta B_{\phi}\mid \gg \mid \delta B_{r} \mid ~{\rm and} 
\mid \delta B_{z}\mid$. This ultimately renders the azimuthally averaged magnetic
fields which is given by $<{\bf{B}}>=\hat{\phi}<B_{\phi}>$ \citep{Oda-etal07}.

\subsection{Model Equations}

In this work, a thin, axisymmetric, magnetized accretion flow onto a rotating
black hole is considered and the accretion disc is assumed to lie on the black hole
equatorial plane. Moreover, we employ the cylindrical polar coordinate ($x,\phi,z$) 
to study the properties of accretion flow, where black hole is placed at its origin.
In order to express the flow variables, we choose
an unit system as $M_{\rm BH}=c=G=1$, where $M_{\rm BH}$ is the mass
of the black hole, $c$ represents the speed of light and $G$ denotes
the gravitational constant, respectively.
Accordingly,  length,  angular momentum and time are measured in units of
$GM_{\rm BH}/c^2$, $GM_{\rm BH}/c$ and $GM_{\rm BH}/c^3$, respectively. 
In the subsequent sections, we choose $M_{\rm BH} = 10M_{\odot}$ as a reference value.

Considering steady state scenario, the governing equations of motion that describe
the magnetized accreting matter are obtained as follows:

\noindent (i) Equation for radial momentum:

$$
{v\frac{dv}{dx} + \frac{1}{\rho}\frac{dP}{dx} 
+ \frac{d\Psi_{\rm eff}}{dx} + \frac{\left<B_{\phi} ^2\right>}{4\pi x \rho} = 0},
\eqno(1)
$$
where $v$ and $\rho$ stand for the radial velocity and density of the
flow and $P$ represents total pressure
which we take into account as $P=p_{\rm gas}+p_{\rm mag}$ where, $p_{\rm gas}$ 
and $p_{\rm mag}$ denote the gas pressure and the magnetic pressure of the flow. 
We obtain the gas pressure inside the disc as $p_{\rm gas} = R \rho T/\mu$, where $R$,
$T$ and $\mu$, respectively, represent the gas constant, the temperature and the
mean molecular weight.
Here, we use $\mu=0.5$ for fully ionized hydrogen. Further, the magnetic pressure
is obtained as
$p_{\rm mag} = <B_{\phi}^2>/8\pi$. We define
$\beta = p_{\rm gas}/p_{\rm mag}$ and using this, we attain total pressure as 
$P = p_{\rm gas} (1+\beta)/\beta$. Moreover, in equation (1), $\Psi_{\rm eff}$ denotes
the effective pseudo potential around a rotating black hole \citep{Chakrabarti-Mondal06}
and is given by,
$$
\Psi_{\rm eff} = - \frac{{\mathcal Q} + \sqrt{{\mathcal Q^2} - 4{\mathcal P}{\mathcal R}}}{2{\mathcal P}},
$$
where
$$
{\mathcal P} = \frac{\epsilon^2\lambda^2}{2x^2},
$$
$$
{\mathcal Q} = -1 + \frac{\epsilon^2 \omega \lambda r^2}{x^2} + \frac{2a_k\lambda}{r^2 x},
$$
$$
{\mathcal R} = 1 - \frac{1}{r - x_0} + \frac{2a_k\omega}{x} + \frac{\epsilon^2 \omega^2 r^4}{2x^2}.
$$
Here, $x$ represents the cylindrical radial distance and $r$ specifies spherical radial distance,
respectively. Also, $\lambda$ stands for the specific angular momentum of the flow. 
In addition, we write $x_0 = 0.04 + 0.97a_k + 0.085a_k^2$, $\omega = 2a_k /(x^3 +a_k^2 x+2a_k^2 )$ and
$\epsilon^2 = (x^2 - 2x + a_k^2 )/(x^2 +a_k^2 +2a_k^2 /x)$, where $\epsilon$ refers the redshift
factor and $a_k$ denotes the spin of the black hole. It is to be noted that 
the adopted pseudo potential satisfactorily mimics
the space-time geometry around rotating black hole for $a_k \lesssim 0.8$
\citep{Chakrabarti-Mondal06}.

\noindent (ii) Mass flux conservation equation:
$$
\dot{M}=2\pi xv\Sigma,
\eqno(2)
$$
where $\dot{M}$ specifies the accretion rate which we treat as global constant all 
throught and $\Sigma$ represents the vertically integrated density
\citep{Matsumoto-etal84}. It may be noted that in this work, the direction of
the inward radial velocity is considered as positive always. 

\noindent (iii) Azimuthal momentum conservation equation:
$$
v\frac{d\lambda(x)}{dx}+\frac{1}{\Sigma x}\frac{d}{dx}(x^2T_{x\phi}) = 0.
\eqno(3)
$$
Here, we assume the vertically integrated total stress to be dominated
by the $x\phi$ component of the Maxwell stress $T_{x\phi}$. For the
accretion flow with large radial velocity, $T_{x\phi}$ comes out to be
 \citep{Chakrabarti-Das04,Machida-etal06}
$$
T_{x\phi} = \frac{<B_{x}B_{\phi}>}{4\pi}h = -\alpha_{B}(W + \Sigma v^2),
\eqno(4)
$$
where $h$,  $\alpha_B$ and $W$, respectively, represent the local disc height,
the proportionality constant and the vertically 
integrated pressure of the flow \citep{Matsumoto-etal84}. Following the work of
\citet{Shakura-Sunyaev73}, we regard $\alpha_B$ as a global constant all throughout
of the flow. Note that when $v$ is significantly small, as in the case of Keplerian disc, 
equation (4) reduces to `$\alpha$-model'  \citep{Shakura-Sunyaev73}.

We consider thin disc approximation where infalling matter maintains 
hydrostatic equilibrium in the vertical direction and calculate the 
disc height ($h$) as,
$h = a \sqrt{x/(\gamma \Psi_{r}^{'})}$
where $\Psi_{r}^{'} = \left(\frac{\partial \Psi_{\rm eff}}{\partial r}\right)_{z << x}$, $z$ 
denotes local vertical scale height in the cylindrical coordinate system and $r =\sqrt {x^2 + z^2}$ 
\citep{Das-etal10}. Here, we define the sound speed as $a=\sqrt {\gamma P/\rho}$,
where $\gamma$ stands for the adiabatic index of the flow.
In this work, we assume $\gamma$ to remain constant along the flow and choose
$\gamma=4/3$. 

\noindent (iv) The equation for entropy:

$$
\Sigma v T \frac {ds}{dx}=\frac{hv}{\gamma-1}
\left(\frac{dp_{\rm gas}}{dx} -\frac{\gamma p_{\rm gas}}{\rho}\frac{d\rho}{dx}\right)=Q^- - Q^+,
\eqno(5)
$$
where $T$ and $s$ refer to the temperature and specific entropy of the flow, respectively. 
Moreover, $Q^+$ denotes the 
heating rate and $Q^-$ represents the 
cooling rate of the flow. Meanwhile, the numerical simulation works of
\citet{Hirose-etal06,Machida-etal06} indicate that during accretion,
heating of the accreting matter occurs because of the energy dissipation
via magnetic reconnection process and is calculated as
$$
Q^{+} = \frac{<B_{x}B_{\phi}>}{4\pi} x h \frac{d\Omega}{dx} = 
-\alpha _{B}(W + \Sigma v^2) x \frac{d\Omega}{dx},
\eqno(6)
$$
where $\Omega$ stands for the angular velocity of the flow.

Usually, the accretion flow experiences heat loss as the consequences
of the variety of cooling mechanisms, such as 
bremsstrahlung, synchrotron and 
Comptonization of bremsstrahlung as well as synchrotron photons. However,
in the present study, as the infalling matter is magnetized in nature,
we therefore consider only the synchrotron radiative mechanism as dominant cooling
process and the corresponding cooling rate is obtained as \citep{Shapiro-Teukolsky83}, 
$$
Q^-= \frac{Sa^5\rho h}{v} \sqrt{\frac{\Psi_{r}^{'}}{x^3}} \frac{\beta^2}{(1+\beta)^3},
\eqno(7)
$$
with,
$$
S= 1.4827 \times 10^{18} \frac{ {\dot m} \mu^2 e^4}{I_n m_e^3\gamma^{5/2}}
\frac{1}{GM_{\odot}c^3},
$$ 
where $e$ and $m_e$ represent the charge and mass of the electron
and $\dot m$ denotes the accretion rate expressed in units of 
Eddington rate ($\dot{M}_{\rm Edd} = 1.39 \times 10^{17} \times M_{\rm BH}/M_{\odot}$ gm~s$^{-1}$).
Also, $I_n = (2^n n!)^2/(2n + 1)!$ and $n$ represents the polytropic
index of the flow which is related to the adiabatic index as $n= 1/(\gamma - 1)$.
We estimate the electron temperature employing the relation 
$T_e = (\sqrt{m_e/m_p})T_p$, where the coupling between ion and electron 
is neglected \citep{Chattopadhyay-Chakrabarti02}. Here, $m_p$ and $T_p$
refer the mass and temperature of the ion. 
Note that in this work, we ignore the bremsstrahlung emission process as it is an inefficient cooling process for stellar mass black hole system \citep{Chattopadhyay-Chakrabarti02}. Moreover, we also disregard the inverse Comptonization process as well although its contribution may not be negligible especially at the inner part of the disc. Nevertheless, we make this assumption simply because the framework of single temperature accretion flow does not allow one to study the Componization process as it requires the consideration of two-temperature flow. However, we infer that when both synchrotron and Compton processes are present, the accretion flow will experience more dissipation and therefore, the results we present in the subsequent sections are expected to modify quantitatively although the overall conclusions perhaps be remain qualitatively unaltered. 

\noindent (v) The advection equation of toroidal magnetic flux:

Following induction equation, the advection rate of toroidal magnetic flux
is obtained as,

$$
\frac {\partial <B_{\phi}>\hat{\phi}}{\partial t} = {\bf \nabla} \times
\left({\vec{v}} \times <B_{\phi}>\hat{\phi} -{\frac{4\pi}{c}}\eta {\vec{j}}\right),
\eqno(8)
$$
where $\vec {v}$, ${\vec{j}}$ and $\eta$, respectively, represent the velocity vector, 
the current density  and the resistivity of the flow. It may be noted that equation 
(8) is azimuthally averaged. For an accretion disc, since the Reynold number is
generally very large, we ignore the magnetic-diffusion terms because of 
large length scale. Furthermore, here we ignore dynamo term as well.
Considering steady state, the obtained equation is further vertically integrated
employing the assumption that the azimuthally averaged toroidal magnetic fields
disappear at disc surface.
Based on these considerations, the toroidal
magnetic flux advection rate is calculated as,
$$
\dot{\Phi} = - \sqrt{4\pi}v h {B}_{0} (x),
\eqno(9)
$$
where
\begin{eqnarray*}
{B}_{0} (x) && = \langle {B}_{\phi} \rangle \left(x; z = 0\right)  \nonumber \\
&& = 2^{5/4}{\pi}^{1/4}(R T/\mu)^{1/2}{\Sigma}^{1/2}h^{-1/2}{\beta}^{-1/2}
\end{eqnarray*}
denotes azimuthally averaged toroidal magnetic field resided at the equatorial plane 
of the accretion disc \citep{Oda-etal07}. Inside the accretion disc, 
if the magnetic flux is dissipated by the magnetic reconnection or escapes ̇
from the disc due to buoyancy, $\dot{\Phi}$ will not be conserved. 
Besides, when MRI driven dynamo augments the toroidal magnetic flux, $\dot{\Phi}$ 
may vary with radial coordinate. Keeping these findings in mind, we thus consider 
$\dot{\Phi} \propto x^{-\zeta}$ \citep{Oda-etal07}, where $\zeta$ 
stands for a parameter describing the magnetic flux advection rate.
Therefore, we have the following parametric relation as
$$
\dot{\Phi}\left(x; \zeta, \dot{M}\right) \equiv \dot{\Phi}_{\rm edge} 
\left(\frac{x}{x_{\rm edge}} \right)^{-\zeta},
\eqno(10)
$$
where $\dot{\Phi}_{\rm edge}$ indicates the advection rate of the toroidal 
magnetic field at a large distance, usually the disc outer edge
($x_{\rm edge}$). For $\zeta = 0$, radial magnetic flux remains conserved
whereas, for $\zeta > 0$, the magnetic flux is increased with the decrease of $x$. 
However, for representation, in this study, we choose $\zeta=1$
all throughout unless stated otherwise.

\subsection{Analysis of transonic conditions}

During the course of accretion, matter from the outer edge of the disc 
($x_{\rm edge}$) proceeds towards the black hole under the influence of gravity.
In reality, inflowing matter possesses negligible radial velocity at $x_{\rm edge}$
in contrast with the local sound speed and enters into the black hole
with velocity equivalent to $c$. This findings evidently demand the transonic
nature of the accreting matter.
The radial coordinate where the accretion flow smoothly changes its sonic
character from subsonic to supersonic state is commonly called as critical
point. In order to analyze the transonic conditions, we simultaneously
solve equations (1), (2), (3), (5), (9) and (10) and obtain the
wind equation \citep[and references therein]{Das07} which is given by,
$$
\frac {dv}{dx}=\frac{N}{D},
\eqno(11)
$$
where the numerator ($N$) is calculated as,

$$
N =\frac{Sa^5}{v}\sqrt{\frac{\Psi_{r}^{'}}{x^3}}\frac{\beta^{2}}{(1+\beta)^{3}} +\frac {2\alpha^2_B I_n  (a^2g+\gamma v^2)^2}{\gamma^2 x v}
$$
$$
+\left[\frac {[3+\beta(\gamma+1)]v}{(\gamma-1)(1+\beta)}
 -\frac{4\alpha^2_B g I_n (a^2g+\gamma v^2)}{\gamma v} \right]
\left(\frac{d\Psi_{\rm eff}}{dx}\right)
$$
$$
+\left[\frac {v a^2(2\beta\gamma+4)}{2\gamma(\gamma-1)(1+\beta)}-
\frac {2\alpha^2_B I_n a^2g(a^2g+\gamma v^2)}{\gamma^2 v}\right]
\left(\frac{d{\rm ln}\Psi_{r}^{'}}{dx}\right)
$$
$$
+\frac {2\{3 + \beta(\gamma + 1)\}a^2 v}{\gamma x (\gamma - 1)(1 + \beta)^2}-\frac {3a^2 v(2\gamma\beta+3)}{2\gamma x (1 + \beta) (\gamma - 1)}
$$
$$
+\frac {6 \alpha_B^2 I_n a^2 g(a^2g+\gamma v^2)}{\gamma^2 v x}-\frac {8 \alpha_B^2 I_n a^2 g(a^2g+\gamma v^2)}{\gamma^2 v (1 + \beta) x}
$$
$$
- \frac {a^2 v (4\zeta - 1)}{2\gamma (1 + \beta)(\gamma - 1)x}-\frac {4\lambda \alpha_B I_n (a^2g+\gamma v^2)}{\gamma x^2} 
\eqno(11a)
$$

and the denominator ($D$) is calculated as,
$$
D = \frac {2a^2(2 + \gamma \beta)}{\gamma(\gamma-1)(1+\beta)} 
-\frac {\{3+\beta(\gamma+1)\}v^2}{(1+\beta)(\gamma-1)}
$$
$$
+\frac{2\alpha^2_B I_n (a^2g+\gamma v^2)}{\gamma}
\left[ (2g-1)-\frac {a^2g}{\gamma v^2}\right].
\eqno(11b)
$$
In the above analysis, we define $g=I_{n+1}/I_{n}$.

Next, we calculate the derivative of $a$, $\lambda$ and $\beta$ with respect to $x$ as,

$$
\frac{da}{dx}= - \left( \frac{\gamma v}{a} - \frac{a}{v}\right)
\frac{dv}{dx} + \frac{3a}{2x} -\frac{a}{2}\left(\frac{d{\rm ln}\Psi_{r}^{'}}{dx}\right) 
$$
$$
- \frac{\gamma}{a}\left(\frac {d\Psi_{\rm eff}}{dx}\right) - \frac{2a}{(1+\beta)x}
\eqno(12)
$$

$$
\frac{d\lambda}{dx}=
-\frac{\alpha_{B} x (a^2g- \gamma v^2)}{\gamma v^2}\frac{dv}{dx}
+\frac{2 \alpha_{B} axg }{\gamma v}\frac{da}{dx}
$$
$$
+\frac{\alpha_{B}(a^2g+\gamma v^2)}{\gamma v}
\eqno(13)
$$

$$
\frac{d\beta}{dx}= \left[\frac{4(1+\beta)}{v}-\frac{3\gamma v(1+\beta)}{a^2}\right]\frac{dv}{dx}+\frac{9(1+\beta)}{2x}
$$
$$
-{2(1+\beta)}\left(\frac{d{\rm ln}\Psi_{r}^{'}}{dx}\right)-\frac{3\gamma(1+\beta)}{a^2}\frac{d\Psi_{\rm eff}}{dx}
$$
$$
-\frac{6}{x} +\frac{(1+\beta)(4\zeta-1)}{2x}
\eqno(14)
$$

Since the accretion solutions must be smooth along the streamline, the radial
velocity gradient ($dv/dx$) will be inevitably real and finite at every radial 
coordinate. Nevertheless, equation (11b) is revealed the fact that between
$x_{\rm edge}$ and the black hole horizon, there is a possibility where the
denominator ($D$) may vanish at some point.
In order for maintaining the flow to become smooth always, 
it is therefore necessary
that the location where $D$ goes to zero, $N$ also must vanish there.
The location where $N$ and $D$ simultaneously disappears has a special significance
and such location is termed as critical point ($x_c$). It is to be noted that accretion flow
becomes transonic at $x_c$ and accordingly, we have two conditions at $x_c$ which 
are obtained by setting $N=0$ and $D=0$, respectively.
Using $D=0$, we calculate the Mach number (defined as the ratio of radial velocity
to the sound speed, $M=v/a$) at $x_c$ as,

$$
M_c =\sqrt {\frac{-m_2 - \sqrt{m^2_2-4m_1 m_3}}{2m_1}},
\eqno(15)
$$
where
$$ 
m_1=2\alpha^2_{B} I_n \gamma^2(\gamma-1)(2g-1)(1 + \beta_c) - \gamma^2\{3 + (\gamma+1)\beta_c\},
$$
$$
m_2=2\gamma(2 + \gamma\beta_c) + 4\alpha^2_{B} I_n \gamma g (g-1)(\gamma-1)(1 + \beta_c),
$$ 
$$
m_3=-2\alpha^2_{B} I_n g^2 (\gamma-1) (1 + \beta_c).
$$

Setting $N=0$, we obtain a cubic equation of sound speed ($a_c$) at $x_c$ as,

$$
{\mathcal A}a_c^3 + {\mathcal B}a_c^2 + {\mathcal C}a_c +{\mathcal D}= 0 ,
\eqno(16)
$$
where
$$
{\mathcal A}=S\sqrt{\frac{\Psi_{r}^{'}}{x_c^3}}\frac{\beta_c^{2}}{(1+\beta_c)^{3}},
$$

$$
{\mathcal B} =  \frac {2\alpha^2_B I_n (g+\gamma M_c^2)^2}
{\gamma^2 x_c}+\frac{M_c^2(2\gamma\beta_c+4)}{2\gamma(\gamma-1)(1+\beta_c)}\left(\frac{d{\rm ln}\Psi_{r}^{'}}{dx}\right) 
$$
$$
-\frac {2\alpha^2_B I_n g(g+\gamma M_c^2)}
{\gamma^2}\left(\frac{d{\rm ln}\Psi_{r}^{'}}{dx}\right)
$$
$$
 + \frac{2\{3+\beta_c(\gamma+1)\}M_c^2}{\gamma x_c (\gamma-1)(1+\beta_c)^2} 
$$
$$
 - \frac{3M_c^2(2\gamma\beta_c+3)}{2\gamma(\gamma-1)(1+\beta_c)x_c}+\frac {6\alpha^2_B I_n g(g+\gamma M_c^2)}{\gamma^2 x_c}
$$
$$
-\frac {8\alpha^2_B I_n g(g+\gamma M_c^2)}{\gamma^2 (1+\beta_c)x_c}- \frac{(4\zeta - 1)M_c^2}{2\gamma(\gamma-1)(1+\beta_c) x_c},
$$

$$
{\mathcal C} = -\frac {4\lambda_c \alpha_B I_n M_c (g+\gamma M_c^2)}{\gamma x_c^2},
$$

$$
{\mathcal D} = \left[\frac {[3+\beta_c(\gamma+1)]M_c^2}{(1+\beta_c)(\gamma-1)}-\frac{4\alpha^2_B g I_n (g+\gamma M_c^2)}{\gamma} \right]
$$
$$
\times\left(\frac{d\Psi_{\rm eff}}{dx}\right).
$$
Here, the flow variables specified using subscript `c' denote their values evaluated at $x_c$.

Now, using the accretion flow parameters, we solve equation (16) to obtain
the sound speed ($a_c$) at $x_c$  and subsequently calculate $v_c$ using equation (15). By employing the
values of $v_c$ and $a_c$ in Eq. (11), we examine the characteristics of the critical 
points. At the critical point, we get $(dv/dx)=0/0$ and thus, we use l$^{'}$Hospital rule
for obtaining the value of $(dv/dx)$ at $x_c$ (hereafter, $(dv/dx)_c$). Usually, 
$(dv/dx)_c$ owns two values; one for accretion and the other for wind.  
When the values of $(dv/dx)_c$ are real and of opposite sign, the
critical point is known as saddle type \citep{Chakrabarti-Das04} and this type
of critical point is particularly important due to the fact that transonic solution
can cross it smoothly.  In the present study, since our motivation is to investigate the
structure of the magnetized accretion flow,  we therefore focus into the accretion
solutions only in the subsequent analysis.

\section{Results and Discussions}

\subsection{Transonic Global Solutions}

In this work, we intend to obtain the global magnetized transonic accretion 
solution that  delineates a smooth connection between horizon and the
disc edge.
With this aim, we simultaneously solve the equations (11-14)
for a specified set of flow parameters. While doing this, we treat 
$\dot{m}$, $\alpha_B$, and $\gamma$ as global parameters of the flow.
Moreover, one requires $a_k$ value and the boundary values of
$\lambda$ and $\beta$ at a given $x$ as local parameters to solve these equations.
Note that we express angular momentum ($\lambda$) in terms of Keplerian
	angular momentum $\lambda_{\rm K}$ ($\equiv\sqrt{x^3/(x-2)^2}$) all throughout the paper.
Since the black hole
accretion solutions are necessarily transonic in nature, flow must pass through
at least one critical point and therefore, it is reasonable to choose  the boundary values of the
flow at the critical point. With this, we hereby integrate equations (11-14) starting
from the critical point once inwards up to just outside the black hole horizon and
then outward up to a large distance (equivalently `outer edge of the disc'). Ultimately,
these two parts of are joined to obtain a complete global transonic accretion solution.
Depending on the input parameters, accretion flow may possess single or multiple
critical points \citep{Das-etal01a,sarkar2013}. These critical points are classified
as inner ($x_{\rm in}$) or outer ($x_{\rm out}$) critical points depending on whether
they form close to or far away from the black hole horizon.

\subsection{Global Accretion Solutions with Shock}

When the accretion flow containing multiple critical points accretes
on to a black hole, it first passes through the outer critical point ($x_{\rm out}$)
to become supersonic and keeps on accreting further inwards.
Meanwhile, flow starts experiencing centrifugal repulsion
resulting the accumulation of matter in the nearby region of the black hole that
ultimately induces the shock transition when the density threshold is reached.
With this, an effective virtual barrier is formed around the black hole. At shock, 
supersonic flow jumps in to the subsonic branch that makes
the post-shock flow hot as the kinetic energy of the flow is converted to the thermal energy. Moreover,
across the shock, flow undergoes shock compression that ultimately causes the post-shock flow to
become dense. Interestingly, 2$^{nd}$ law of thermodynamics suggests that shocked accretion
solutions are favorable as the entropy of the post-shock matter is comparatively higher
than the pre-shock matter \citep{Becker-Kazanas01}.
We calculate the entropy of the flow which
is expressed as \citep{Chakrabarti96a},
$\dot{\cal{M}}(x)=vxa^{2n+1}\left(\frac{\beta}{1 + \beta}\right)^n \sqrt{\frac{x}{\gamma\Psi_{r}^{'}}}$.
In the dissipation free limit, $\dot{\cal{M}}$ remains constant all throughout expect at the
shock transition.
What is more is that at the discontinuous transition, the conservation of mass flux, momentum
flux, energy flux and magnetic flux are held in order to satisfy the standing shock
conditions
\citep[and reference therein]{sarkar2016} and hence, these conservation laws across the
shock front can be written as the continuity of (a) mass flux (${\dot M}_{-}={\dot M_{+}}$)
(b) the momentum flux  ($W_{-}+\Sigma_{-} \upsilon^2_{-}=W_{+}+\Sigma_{+} \upsilon^2_{+}$)
(c) the energy flux (${\cal E_{-}}={\cal E_{+}}$) and (d) the magnetic flux ($\dot {\Phi}_{-}=\dot {\Phi}_{+}$),
respectively. In this work, we consider the shock to be thin and non-dissipative and the 
flow variables with subscripts `$-$' and `$+$' represent their values just before and after the shock.
Following \citet{Fukue90, Samadi-etal14}, we calculate the local energy of the magnetized dissipative
accretion flow as
${\cal E}(x) = \upsilon^2/2+a^2/(\gamma-1)+\Psi_{\rm eff} 
+<B_\phi ^2>/(4\pi \rho)$,
where all the above quantities bear their usual meaning. In the subsequent analysis, upon employing
the above set of shock conditions, we compute the shock position and its diverse properties knowing the input parameters of the accretion flow.

\begin{figure}
\begin{center}
\includegraphics[width=0.45\textwidth]{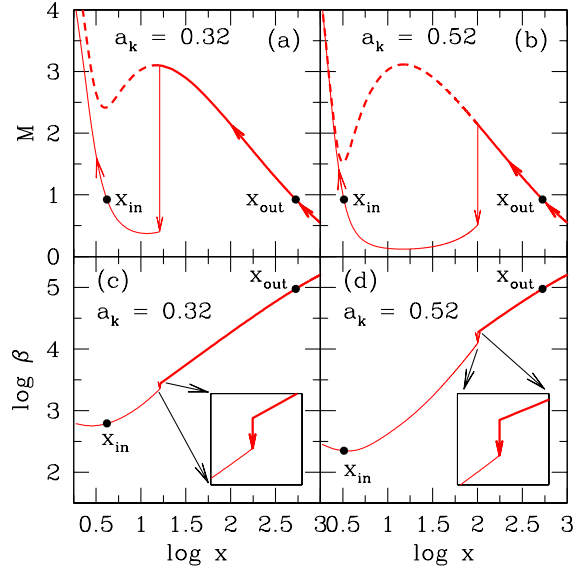}
\end{center}
\caption{(a-b) Plot of Mach number with logarithmic radial distance. 
Flow is injected with $x_{\rm edge} = 1000$, $\lambda_{\rm edge}=0.124\lambda_{\rm K}$, 
${\cal E}_{\rm edge}=1.0793 \times 10^{-3}$, $\beta_{\rm edge} = 1.6 \times 10^5$, 
$\alpha_B = 0.02$ and $\dot{m} = 0.05$, respectively. We choose $a_k = 0.32$ in
panel (a) and $a_k = 0.52$ in panel (b). (c-d) Logarithmic variation of plasma-$\beta$ corresponding to solutions (a) and (b). In each panel, $x_{\rm in}$ and
$x_{\rm out}$ are indicated using filled circles and shock transition
is shown by vertical arrow. See text for details.}
\end{figure}

In Fig. 1, we show the result obtained from one representative case where the variation of Mach number ($M = \upsilon/a$) with the logarithmic  radial distance is depicted. 
We choose the injection parameters of the flow at the outer edge ($x_{\rm edge} = 1000$) as 
${\cal E}_{\rm edge}=1.0793 \times 10^{-3}$, $\beta_{\rm edge} = 1.6 \times 10^5$, $\alpha_B = 0.02$ and $\dot{m} = 0.05$, respectively. 
In Fig. 1(a), we consider the black hole to be slowly rotating having $a_k = 0.32$ and
the flow is injected with angular momentum, $\lambda_{\rm edge}=0.124\lambda_{\rm K}$. Here,
flow is subsonic at the outer edge and becomes supersonic after crossing the outer
critical point located at $x_{\rm out} = 530.90$. The supersonic flow proceeds further
inwards and encounters shock transition at $x_s = 16.20$ while jumping in to the
subsonic branch. In the figure, shock position is shown using vertical
arrow. Gradually flow velocity is increased as it moves inward and then it passes 
$x_{\rm in}$ smoothly at $4.1777$ before crossing the horizon.
Here, we show the direction of the flow motion using arrows and mark the inner
and outer critical points with filled circles.
Next, we intend to examine the role of black hole spin in deciding the shock transition and hence we inject matter on to a moderately rotating black hole ($a_k = 0.52$) keeping all the flow parameters same as in Fig. 1a. It may be noted that for this chosen set of flow parameters, standing shock solution ceases to exist when $a_k > 0.52$.
The result is depicted in Fig. 1b, where the outer critical point, shock location
and inner critical point are obtained as $x_{\rm out} = 531.43$, $x_{\rm s} = 100.62$ and
$x_{\rm in} = 3.2502$, respectively. Since all the flow parameters at the outer edge of the disc are kept fixed including angular momentum, 
we observe that shock forms at larger radial distance for $a_k=0.52$. In reality, a 
spinning black hole distorts the space-time fabric in its vicinity, allowing matter to orbit at a closer distance as compared to a non-rotating one. Due to the effect of frame dragging, the fluid angular momentum is affected by the rotation of the black hole. It is known that the shock formation in accretion flow happens as a result of the competition between the gravitational pull and the centrifugal repulsion. When flow is injected from the outer edge of the disc with fixed boundary conditions, because of the spin-orbit coupling term in the Kerr geometry, the increase of spin parameter ($a_k$) modifies the angular momentum profile of the flow and the shock front is pushed away from the horizon as is observed in Fig. 1. This finding is consistent with the results of \citet{Aktar-etal15}. 
Overall, we see that the standing shock in magnetized flow is continue to
exist around the rotating black hole and  when $a_k$ is increased, shock transition occurs for relatively low angular momentum flow and {\it vice versa}. Further, in panel (c) and (d), we show the variation of plasma-$\beta$ with $\log~x$ corresponding to solutions presented in panel (a) and (b), respectively. In both the cases, we find that plasma-$\beta$ steadily decreases with the decrease of radial coordinate. This clearly indicates that the magnetic activity inside the disc increases as the flow accretes towards the horizon.

\subsection{Properties of Standing Shocks}

\begin{figure}
\begin{center}
\includegraphics[width=0.45\textwidth]{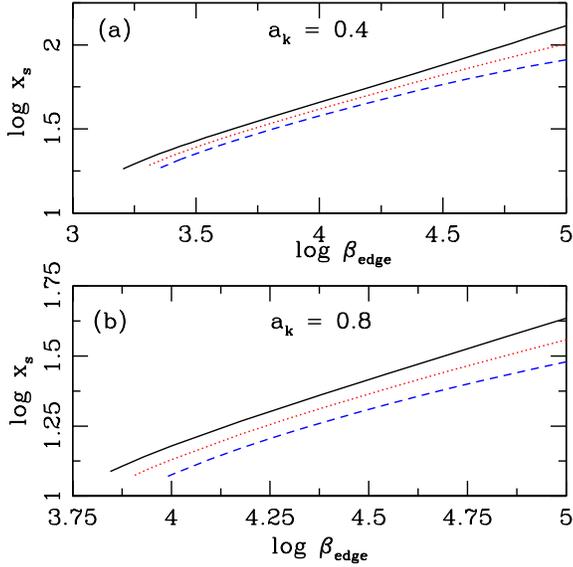}
\end{center}
\caption{Shock location ($x_s$) variation with $\beta_{\rm edge}$. Here, the inflow parameters are chosen as $x_{\rm edge} = 1000$, ${\cal E}_{\rm edge} = 1.0793 \times 10^{-3}$, $\alpha_B = 0.02$ and $\dot{m} = 0.05$, respectively. In every panel, spin of the black hole ($a_k$) is marked. In panel (a), results plotted with solid, dotted and dashed curves are obtained for $\lambda_{\rm edge} = 0.12845\lambda_{\rm K}, 0.12791\lambda_{\rm K}$ and $0.12737\lambda_{\rm K}$. And in panel (b), results depicted with solid, dotted and dashed curves are for $\lambda_{\rm edge} = 0.11443\lambda_{\rm K}, 0.11390\lambda_{\rm K}$ and $0.11336\lambda_{\rm K}$, respectively. See text for details.
}
\end{figure}

One of the pertinent aspect in understanding the magnetically supported
accreting flow around the rotating black holes is to study the dependence
of the shock position $(x_s)$ on the $\beta$ values. Accordingly, we calculate $x_s$
in terms of $\beta_{\rm edge}$ for flows with fixed outer boundary values
accreting on to a given black hole. 
For that, we choose the outer boundary
parameters as $x_{\rm edge} = 1000$, ${\cal E}_{\rm edge} = 1.0793 \times 10^{-3}$,
$\alpha_B = 0.02$ and $\dot{m} = 0.05$. In Fig. 2(a), we display the results
obtained for $a_k = 0.4$, where solid, dotted and dashed curves are for $\lambda_{\rm edge} = 0.12845\lambda_{\rm K}$, $0.12791\lambda_{\rm K}$ and $0.12737\lambda_{\rm K}$, respectively. We notice
that the shock front proceeds towards the horizon with the decrease of
$\beta_{\rm edge}$ irrespective to the values of $\lambda_{\rm edge}$.
This happens because when $\beta_{\rm edge}$ is decreased, the efficiency
of synchrotron cooling is enhanced due to the increase of magnetic activity
inside the disc. The effect becomes more prominent at the inner part of the
disc ($i.e.$, PSC) as, due to shock transition, both density and temperature
are relatively higher there compared to the pre-shock flow. This renders the
thermal pressure to drop down in the PSC region and ultimately shock front
moves inward to maintain the pressure balance across it.
Incidentally, keeping the all the boundary flow parameters fixed, one can
not reduce $\beta_{\rm edge}$ indefinitely as shock ceases to exist
when $\beta_{\rm edge} < \beta_{\rm edge}^{\rm cri}$ (shock conditions
fail to satisfy there). It may be noted that $\beta_{\rm edge}^{\rm cri}$
does not have a universal value, instead it depends on the flow parameters
fixed at the outer edge of the disc. 
Further, we depict the results for $a_k = 0.8$ in Fig. 2b, where solid, 
dotted and dashed curves represent results corresponding to $\lambda_{\rm edge} = 0.11443\lambda_{\rm K}$, $0.11390\lambda_{\rm K}$ and $0.11336\lambda_{\rm K}$, respectively.
Here, we keep all the other flow parameters same as in Fig. 2a.
We find that the shock location proceeds towards the horizon
with the decrease of $\beta_{\rm edge}$ in all cases as observed
in Fig. 2a.

\begin{figure}
\begin{center}
\includegraphics[width=0.45\textwidth]{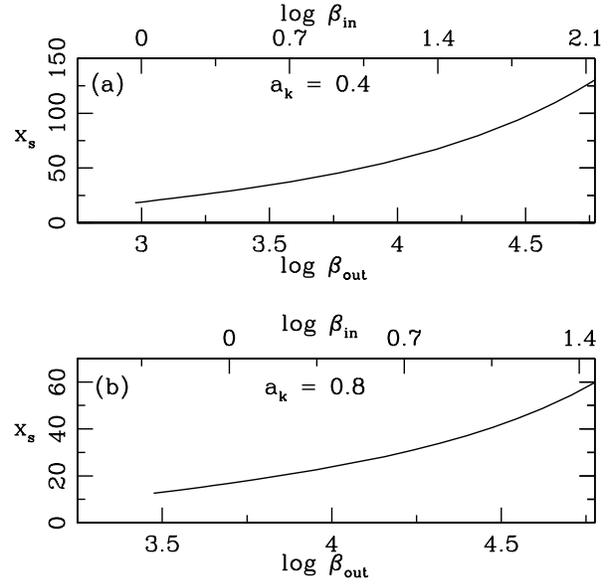}
\end{center}
\caption{Variation of the shock location ($x_s$) as function of $\beta_{\rm out}$
(lower axis) and $\beta_{\rm in}$ (upper axis). In each panel, $a_k$ is marked.
See text for details.
}
\end{figure}

Next, we examine the correlation of $\beta$ values between the inner and outer critical
points for shock induced global accretion solutions. While doing this, we choose  two
cases where inflowing matters are accreted on to rotating black holes having
different spin parameters as $a_k = 0.4$ and $0.8$, respectively. For $a_k = 0.4$, we 
consider the result depicted in Fig. 3a corresponding to $\lambda_{\rm edge} = 0.12845\lambda_{\rm K}$ and show
the variation of shock location as function of both $\beta_{\rm out}$ (lower horizontal axis)
and $\beta_{\rm in}$ (upper horizontal axis).  The other flow parameters are considered same as in Fig. 2. Here, $\beta_{\rm in}$ and $\beta_{\rm out}$
refer $\beta$ values measured at $x_{\rm in}$ and $x_{\rm out}$, respectively. We see that $x_s$ 
decreases when the magnetic activity is increased ($\beta$ is decreased) inside the disc.
We continue our study choosing the result presented in Fig. 3b for $\lambda_{\rm edge} = 0.11551\lambda_{\rm K}$
and show the variation of $x_s$ in terms of $\beta_{\rm out}$ as well as $\beta_{\rm in}$
in Fig. 3b. We observe that in all cases, $\beta_{\rm in} < \beta_{\rm out}$
all throughout. This finding is not surprising because, in our model, the advection of magnetic
flux increases as the inflowing matter approaches towards the horizon and eventually,
$\beta$ is reduced towards the inner part of the disc. Moreover, we find that shock
solutions exist even for $\beta_{\rm in}<1$ irrespective to the choice of $a_k$ value.
This evidently indicates that global transonic accretion solutions harbor standing
shock waves both in gas pressure dominated as well as in magnetic pressure dominated
flows for a wide range of $a_k$ values.

\begin{figure}
\begin{center}
\includegraphics[width=0.45\textwidth]{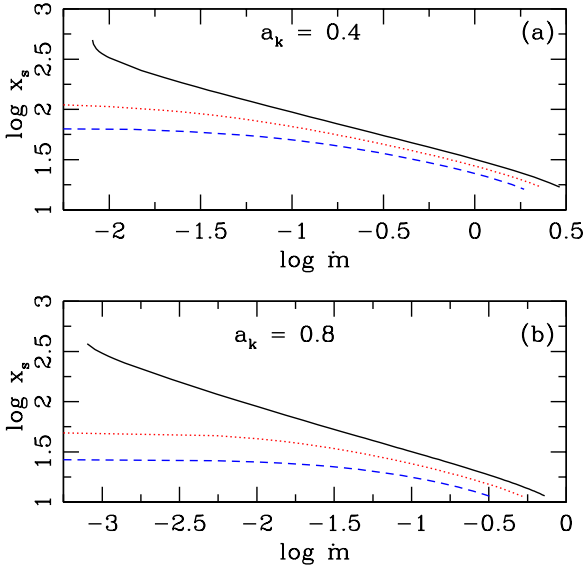}
\end{center}
\caption{Variation of the shock location ($x_s$) as function of $\dot{m}$. Flow
parameters at the outer edge of the disc is chosen as $x_{\rm edge} = 1000$, 
${\cal E}_{\rm edge} = 1.0793 \times 10^{-3}$, $\alpha_B = 0.02$ and 
$\beta_{\rm edge} = 10^5$, respectively. Results depicted in top and
bottom panels are for $a_k=0.4$ and $0.8$.  In (a),  solid, dotted and
dashed curves are obtained for $\lambda_{\rm edge} = 0.12845\lambda_{\rm K}, 0.12737\lambda_{\rm K}$ and $0.12630\lambda_{\rm K}$ whereas in (b), solid, dotted and dashed curves represents results for $\lambda_{\rm edge} = 0.11443\lambda_{\rm K}, 0.11336\lambda_{\rm K}$ and $0.11229\lambda_{\rm K}$. See text for details.
}
\end{figure}

It is worthy to explore the effect of cooling on the formation of shock 
wave in an accretion flow and therefore, in Fig. 4, we study the variation
of shock location ($x_s$) with accretion rate ($\dot{m}$). Towards this,
we consider the flow injection parameters as 
$x_{\rm edge} = 1000$, $\beta_{\rm edge} = 10^5$,
${\cal E}_{\rm edge} = 1.0793 \times 10^{-3}$ and $\alpha_B = 0.02$,
respectively. As before, in Fig. 4a, we chose $a_k = 0.4$ and the profile
of shock location ($x_s$) is presented for various values of 
$\lambda_{\rm edge}$. Here, solid, dotted
and dashed curves represent flows injected with 
$\lambda_{\rm edge} = 0.12845\lambda_{\rm K}, 0.12737\lambda_{\rm K}$ and $0.12630\lambda_{\rm K}$, respectively. From
the figure, it is clear that large range of  $\dot{m}$ admits standing 
shock in magnetized accretion flow.
Moreover, we find that for a given $\lambda_{\rm edge}$,
$x_s$ moves inwards as $\dot{m}$ is increased. In reality, enhanced accretion 
rate boosts the efficiency of
the radiative cooling that causes the flow to lose energy during accretion.
Since PSC is hot and dense, the effect of cooling at PSC becomes profound 
that evidently decreases the post-shock thermal pressure. Consequently, this
compels the shock front to settle down at some smaller distance to fulfill the shock
conditions. Unfortunately, $\dot{m}$ can not be increased indefinitely 
due to the fact that when $\dot{m}$ exceeds its critical value ($\dot{m}^{\rm cri}$),
standing shocks are no longer feasible as the shock conditions fail to satisfy there.
Clearly, $\dot{m}^{\rm cri}$ does not retain a global value, rather it depends
on the flow parameters. It is also apparent that the possibility of standing
shock formation reduces with the increase of $\dot{m}$. Furthermore, it is
intriguing to understand what happens to the flow when standing shock
conditions fail to satisfy. Interestingly, in that case, inner part of the
accretion flow may start to modulate exhibiting the feature of
oscillatory shock \citep[and references therein]{Das-Aktar2015}. Unfortunately, the investigation
of non-steady shock properties is beyond the scope of the present paper.
In addition, we find that for a given $\dot{m}$, shock front recedes away from the black hole
when $\lambda_{\rm edge}$ is increased. 
In reality, the discontinuous shock transition is essentially the manifestation of the competition between centrifugal repulsion and gravity. When $\lambda_{\rm edge}$ is higher, 
accretion flow possesses higher angular momentum that causes the enhanced centrifugal repulsion against gravity. Because of this, shock front is pushed further out when $\lambda_{\rm edge}$ is increased.
This findings establishes the fact that
shocks are centrifugally driven.  In Fig. 4(b), we present the result corresponding to
$a_k = 0.8$,  where solid, dotted and dashed curves represent results obtained for
$\lambda_{\rm edge} = 0.11443\lambda_{\rm K}, 0.11336\lambda_{\rm K}$ and $0.11229\lambda_{\rm K}$, respectively. Here also, we observe
that the formation of shock and its dependence on $\dot{m}$ and $\lambda_{\rm edge}$
are in general similar to the results shown in Fig. 4(a).

\begin{figure}
\begin{center}
\includegraphics[width=0.45\textwidth]{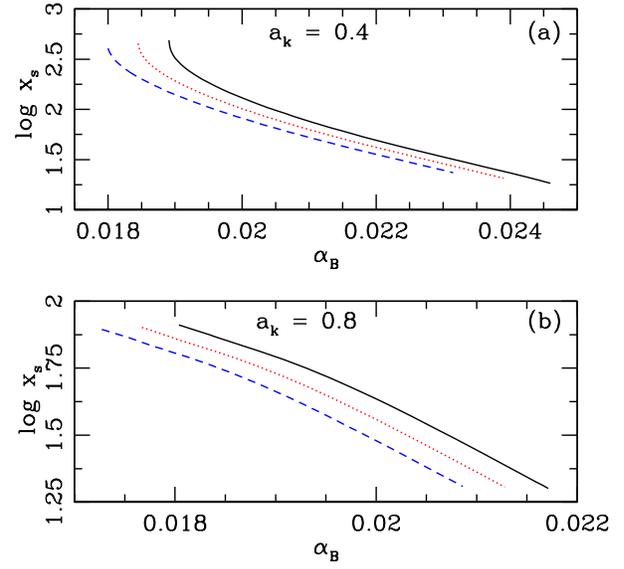}
\end{center}
\caption{Variation of the shock location ($x_s$) as function of $\alpha_B$. Accreting
matter is supplied with inflow parameters as $x_{\rm edge} = 1000$, with ${\cal E}_{\rm edge} = 1.0793 
\times 10^{-3}$, ${\dot m} = 0.05$ and $\beta_{\rm edge} = 10^5$, respectively. In each panel,
$a_k$ is marked. In top panel (a), the results corresponding to $\lambda_{\rm edge} = 0.12845\lambda_{\rm K}, 0.12791\lambda_{\rm K}$ and $0.12737\lambda_{\rm K}$
are represented using solid, dotted and dashed line style. The same line style is used to
denote the results for $\lambda_{\rm edge} = 0.11443\lambda_{\rm K}, 0.11390\lambda_{\rm K}$ and $0.11336\lambda_{\rm K}$ in lower panel (b). See
text for details.
}
\end{figure}

For completeness, we investigate the variation of shock location in terms of
viscosity ($\alpha_B$) for flows having fixed outer edge boundary parameters. 
Here, we choose the flow injection parameters as $x_{\rm edge} = 1000$, ${\cal E}_{\rm edge} = 1.0793 
\times 10^{-3}$, $\beta_{\rm edge} = 10^5$ and $\dot{m} = 0.05$, respectively.
In Fig. 5a, we show the obtained results for $a_k = 0.4$, where solid, dotted and dashed curves are for $\lambda_{\rm edge} = 0.12845\lambda_{\rm K}, 0.12791\lambda_{\rm K}$ and $0.12737\lambda_{\rm K}$, respectively. Notice that shocked accretion solutions
exist for a wide range of $\alpha_B$ and shock location shifts towards the horizon with the increase
of $\alpha_B$ for all cases having different $\lambda_{\rm edge}$ values. In reality, as $\alpha_B$ is increased,
angular momentum transport in the outward direction becomes more efficient
that causes the weakening of centrifugal repulsion and hence, shock front is driven inward.
When $\alpha_B$ exceeds its critical limit ($\alpha_B^{\rm cri}$), shock conditions do not 
remain favorable and as a result, standing shock disappears. Again, it may be noted that
$\alpha_B^{\rm cri}$ largely depends on the accretion flow parameters. Further, in Fig. 5b,
we display the result for  $a_k = 0.8$, where solid, dotted and dashed curves denote results
for $\lambda_{\rm edge} = 0.11443\lambda_{\rm K}, 0.11390\lambda_{\rm K}$ and $0.11336\lambda_{\rm K}$, respectively. Here, we find that shock
location decreases with the increase of $\alpha_B$ around rotating black holes
as well.

\subsection{Parameter Space for Shock}

\begin{figure}
\begin{center}
\includegraphics[width=0.45\textwidth]{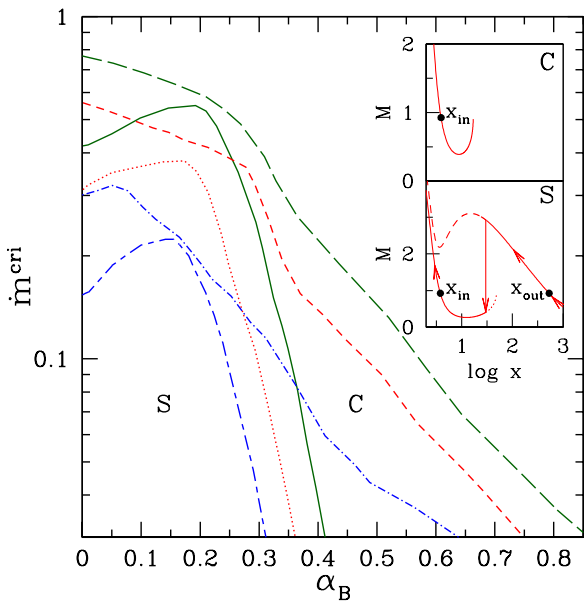}
\end{center}
\caption{Variation of critical accretion rate ($\dot{m}^{\rm cri}$) as a function of
viscosity parameter ($\alpha_B$) for various $a_k$. Here, we choose $\beta_{\rm in} = 10$. Long-dashed,
dashed and dot-dashed curves are obtained for $a_k = 0, 0.4$ and $0.8$ and the region 
bounded by them in $\alpha_B - \dot{m}^{\rm cri}$ plane provides closed accretion
solutions passing through the inner sonic point. In addition, solid, dotted and 
short-long-dashed curves represent the effective region corresponding to
$a_k = 0, 0.4$ and $0.8$ that admits standing shock solutions. In the inset,
examples of closed (marked with C) and shocked solutions (marked with S) 
are presented. See text for details.
}
\end{figure}

We have already mentioned that during the course of accretion, inflowing matter may contain
shock wave provided it possesses multiple critical points. Interestingly, one can obtain standing
shock solution, if the standing shock conditions are satisfied (see \S 3.2). But, when shock conditions
are not favorable and the entropy content at the inner critical point is higher than the outer
critical point, the shock formation never remains steady as the shock location becomes
imaginary \citep{Das-etal01a} and therefore, shock starts to execute continuous back and
forth movements that seems to exhibit the quasi-periodic oscillation phenomenon \citep{Das-etal01a}.
In this case, accretion solution passing through the inner critical point fails to connect the
black hole horizon to the outer edge of the disc as it becomes closed in the range
$x_{\rm in} < x < x_{\rm out}$ with $M (x) = M_c$ \citep{Chakrabarti-Das04}. Needless to mention
that it is not possible to examine the characteristics of the non-steady shock solution in the
framework of the present paper, however, we estimate the critical accretion rate
(${\dot m}^{\rm cri}$) that provides accretion solutions containing standing shocks and/or
closed topologies.
While doing this, we fix $\beta_{\rm in} = 10$, and for a given $a_k$,
we calculate ${\dot m}^{\rm cri}$ as function of $\alpha_B$, where $x_{\rm in}$ and $\lambda_{\rm in}$
are varied freely. Accordingly, in Fig. 6, we classify the parameter space spanned
by $\alpha_B$ and $\dot{m}^{\rm cri}$ that provides closed topologies and standing
shocks, respectively. Examples of closed topology (marked as C) and standing
shock solution (marked as S) are displayed in the small boxes, where the variation of Mach
number with radial coordinate is plotted. In the figure, long-dashed, short-dashed and dot-dashed
curves are obtained for $a_k = 0$, $0.4$ and $0.8$ that separate the $\alpha_B - \dot{m}^{\rm cri}$
plane where left-bottom region allows closed topologies. Similarly, solid, dotted and
short-long-dashed curves separate the standing shock parameter space for $a_k = 0$,
$0.4$ and $0.8$, respectively. We observe that the shock parameter space appears to
be the subset of parameter space for closed topology all throughout. This is expected as
the region of closed topologies includes the region of standing as well as oscillating shocks.
Meanwhile, \citet{Das-Chakrabarti08} showed that for fixed $a_k$, the effective region
of standing shock parameter space shrinks with the increase of accretion rate for an
inviscid flow. Actually, when the accretion rate is enhanced, cooling becomes more effective and
hence, inflowing matter loses energy during accretion. On the other hand, 
viscosity enhances the flow energy as it accretes due to viscous heating. Interestingly, when both dissipation
processes, namely, viscosity and synchrotron cooling, are present in the flow, viscous
dissipation effectively compensates a part of the energy loss happens due to cooling.
Here, in a way, viscosity and cooling act oppositely in deciding the shock parameter
space. However, as synchrotron cooling and viscous heating depend differently on the
flow variables, one does not cancel the other effect completely \citep{Das07}. Overall, for a given
$a_k$, standing shock continues to form
until an optimum combination of ($\alpha_B, \dot{m}^{\rm cri}$) is reached which
evidently exhibits as a peak in the $\alpha_B-\dot{m}^{\rm cri}$ plane.
In general, the flow is dominated by cooling in the left side of the peak whereas viscous
heating dominates on the other side. As expected,
shock disappears when viscosity exceeds its critical limit \citep{Chakrabarti-Das04}.
In addition, in case of a rapidly rotating black hole, shock forms in a relatively low
angular momentum accretion flow \citep{Aktar-etal15} that effectively causes the
weak centrifugal repulsion and therefore, standing shock settles down at a smaller
length scale. Moreover, when the level of dissipation is increased (namely, the
increase of $\alpha_B$ and $\dot{m}$), shock front is compelled to move towards
the horizon (see Fig. 4-5). This clearly indicates that rapidly rotating black holes can
sustain shocks for lower dissipation rates and we observe the similar findings in Fig. 6.

\begin{figure}
\begin{center}
\includegraphics[width=0.45\textwidth]{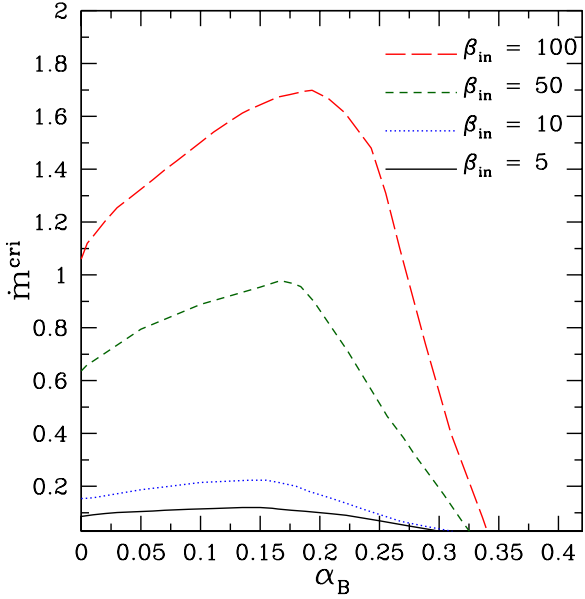}
\end{center}
\caption{Variation of critical accretion rate ($\dot{m}^{\rm cri}$) for standing accretion
shock with viscosity parameter ($\alpha_B$) for different $\beta_{\rm in}$. Here, we fix
black hole spin as $a_k = 0.8$. Solid, dotted, dashed and long-dashed curves denote
results for $\beta_{\rm in} = 5, 10, 50$, and $100$, respectively. See text for details.
}
\end{figure}

Now, we intend to study the effect of magnetic fields in deciding the effective region of
parameter space in ($\alpha_B, \dot{m}^{\rm cri}$) plane for standing shock. In Fig. 7,
we present the obtained results, where shock parameter space is computed for rapidly
rotating black hole ($a_k = 0.8$) considering different $\beta_{\rm in}$ values.
In the figure, the regions bounded with solid, dotted, short-dashed and long-dashed
curves are obtained for $\beta_{\rm in} = 5$, $10$, $50$ and $100$, respectively.
We observe that the effective region of the parameter space for shock gradually
reduces with the decrease of
$\beta_{\rm in}$. This happens due to the fact that when $\beta_{\rm in}$ is low,
synchrotron cooling becomes very much effective and therefore, the level of
dissipation experienced by the inflowing matter turns out to be significant
even with moderate accretion rates.  Thus, the possibility of shock formation
is eventually reduced as the magnetic activity is increased inside the disc.

\begin{figure}
\begin{center}
\includegraphics[width=0.45\textwidth]{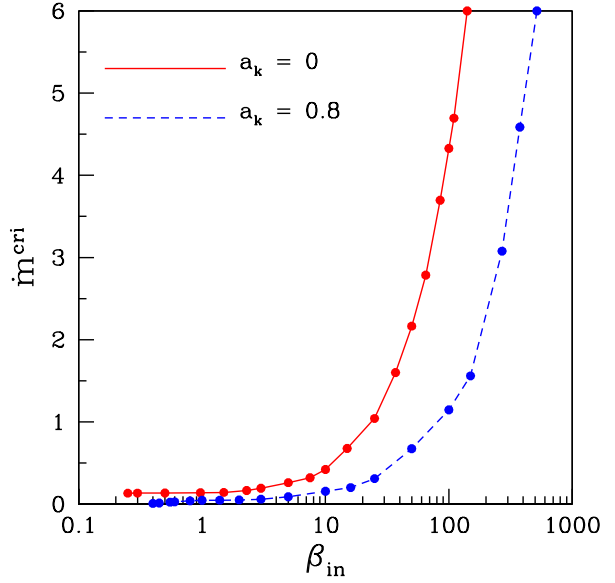}
\end{center}
\caption{Comparison of critical accretion rate $\dot{m}^{\rm cri}$ for shock
with $\beta_{\rm in}$. In the plot, filled circles joined with solid line denote
results for $a_k = 0$ and filled circles connected with dashed line
represent results corresponding to $a_k = 0.8$, respectively. Here 
$\alpha_B = 0.01$ is used. See text for details.
}
\end{figure}

We carry out the analysis further to calculate the critical accretion rate ($\dot{m}^{\rm cri}$)
of the flow as function of $\beta_{\rm in}$ that  provides global accretion
solutions containing standing shock. In Fig. 8, we compare the critical accretion
rate ($\dot{m}^{\rm cri}$) where solid and dashed curves represent the results
obtained for non-rotating ($a_k = 0$) and rapidly rotating ($a_k = 0.8$) black holes,
respectively. Here, we choose the viscosity parameter as $\alpha_B = 0.01$.
We find that standing shocks exist for a wide range of $\beta_{\rm in}$
that effectively includes both gas pressure dominated flows ($\beta > 1$) as well as
magnetic pressure dominated flows ($\beta < 1$). Since synchrotron process
directly depends on the density and magnetic fields of the flow, one can achieve
the desired cooling efficiency by suitably adjusting the accretion rate and plasma
$\beta$. In the figure, we observe this findings for both the cases (for $a_k= 0$
and $0.8$) where the critical accretion rate ($\dot{m}^{\rm cri}$) for shock is found 
to be increased with $\beta_{\rm in}$. In reality, when $\beta_{\rm in} < 1$, the inner part of the
disc is magnetically dominated and a tiny amount of accretion rate is sufficient
to cool the flow. On the other hand, as $\beta_{\rm in}$ is gradually increased,
the strength of magnetic fields becomes weak and therefore, enhanced accretion
rate is needed for the cooling of the flow. Interestingly, when $\beta_{\rm in} \gg 1$, magnetic
fields becomes insignificant and flow is capable of sustaining standing shocks
even for super-Eddington accretion rates ($\dot{m}^{\rm cri} > 1$).
Moreover, we find that for a given $\beta_{\rm in}$, $\dot{m}^{\rm cri}$ is smaller for
higher $a_k$. This clearly indicates that inflowing matter around rapidly rotating
black holes contain shocks for relatively lower accretion rates which is consistent
with the findings of Fig. 6.

\begin{figure}
\begin{center}
\includegraphics[width=0.45\textwidth]{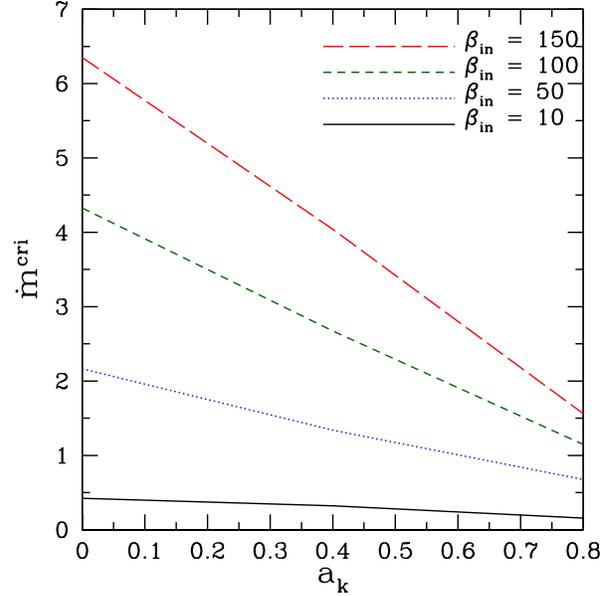}
\end{center}
\caption{Variation of critical accretion rate $\dot{m}^{\rm cri}$ with $a_k$ for shock.
Here, we fix viscosity parameter as $\alpha_B = 0.01$. Results depicted with solid,
dotted, dashed and big-dashed line style correspond to $\beta_{\rm in} = 10, 50, 100$ and $150$.
See text for details.
}
\end{figure}

In the context of the formation of standing shock in an magnetized accretion flow,
we now illustrate the dependence of the critical accretion rate ($\dot{m}^{\rm cri}$) on the
spin of the black hole ($a_k$) in Fig. 9. In order for that we fix the viscosity as
$\alpha_B = 0.01$. Here, solid, dotted, dashed and long-dashed curves are obtained for
$\beta_{\rm in} = 10$, $50$, $100$ and $150$, respectively.  We observe that
for a given $\beta_{\rm in}$, $\dot{m}^{\rm cri}$ decreases with the increase
of $a_k$ in all cases. Moreover, here again we find 
that when $\beta_{\rm in}$ is large, accretion flow continues to sustain standing
shock for higher accretion rate and vice versa. 

\begin{figure}
\begin{center}
\includegraphics[width=0.45\textwidth]{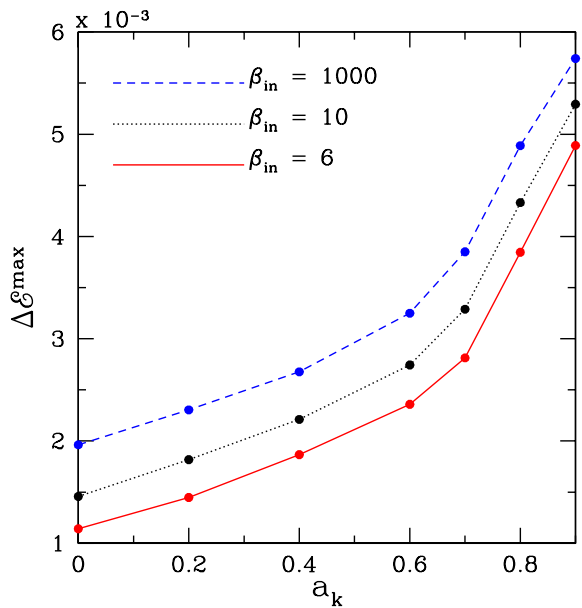}
\end{center}
\caption{Plot of maximum energy dissipation (${\Delta{\cal E}}^{\rm max}$) at the shock 
with $a_k$ for three distinct values of $\beta_{\rm in}$. Here, we choose accretion rate as 
$\dot{m} = 0.05$ and fix viscosity parameter as $\alpha_B = 0.01$. Solid, dotted and
dashed curves are obtained for $\beta_{\rm in} = 6, 10$ and $1000$, respectively. 
See text for details.
}
\end{figure}

\subsection{Energy Extraction from PSC}

So far, we have carried out the investigation of standing shock properties for flows
accreting on to rotating black holes.
While doing this, we consider the shock to be thin and non-dissipative
and therefore, the specific energy remains essentially conserved across the shock front
\citep{Chakrabarti89}. However, in reality, the nature of the shock can be dissipative
as well and in that case, the available energy dissipated at shock escaped through the
disc surfaces along the vertical direction. A part of this energy is then converted to
hard radiations and the rest may be used in jet generation as jets seem to be originated
from the PSC around rotating black holes \citep[and references therein]{Aktar-etal17}.
In effect, this cause the depletion of energy
at PSC \citep{Singh-Chakrabarti11}. Moreover, \citet{Chakrabarti-Titarchuk95} pointed
out that the dissipative energy at shock is likely to be regulated via thermal
Comptonization process that ultimately reduces the thermal energy of the PSC.
Based on the above insight, we model the dissipated energy to be proportional
to the temperature difference between the immediate pre-shock and post-shock
flow. Following this, the energy loss ($\Delta \mathcal{E}$) at the shock is
estimated as \citep{Das-etal10},
$$ 
\Delta \mathcal{E} = fn (a_{+}^{2} - a_{-}^{2}),
\eqno(17)
$$
where $a_{-}$ and $a_{+}$ specify the sound speed just before and after the
shock transition. Here, $f$ refers the fractional value of thermal energy difference
dissipated at shock and we treat it as free parameter
\citep{Das-etal10,sarkar2013,kc13,sarkar2018}. For the
purpose of representation, in this work, we choose $f = 0.1$ all throughout.

In Fig. 10, we show how the maximum energy dissipated at shock 
($\Delta{\cal E}^{\rm max}$) is varied with $a_k$. While doing this, we
choose $\dot{m} = 0.05$ and $\alpha_B = 0.01$, respectively and freely
vary the other flow parameters. In the plot, solid, dotted and dashed curves
illustrate the results for $\beta_{\rm in} = 6, 10$ and $1000$,
respectively. We find that for given $\beta_{\rm in}$, $\Delta{\cal E}^{max}$ 
increases with the increase of $a_k$. In general, 
standing shock forms at a smaller radial coordinate when $a_k$ is
increased \citep{Aktar-etal15} and hence, the thermal energy content
across the shock is also increased. Eventually, the accessible thermal energy
likely to be dissipated at shock is also enhanced. Therefore, for a given
$\beta_{\rm in}$, we find a positive correlation between $\Delta{\cal E}^{\rm max}$
and $a_k$. On the other hand, as $\beta_{\rm in}$ is reduced, 
synchrotron cooling turns out to be more compelling in the flow due to
the increase of magnetic field strength that ultimately reduces the 
thermal energy content in the PSC. Thus, $\Delta{\cal E}^{\rm max}$
diminishes with the decrease of $\beta_{\rm in}$ for fixed $a_k$.
Finally, if the mass, spin and accretion rate of a given black hole candidate
is known, the above formalism can be employed to estimate the maximum
accessible energy in the PSC region and then this unbound energy could be
compared with the observed radio jet kinetic power. Such a task is under
progress and would be reported elsewhere.

\section{Summary}

In this paper, we study the magnetized advection accretion flow around
rotating black hole where viscosity and synchrotron cooling is considered as the
dominant dissipation processes. We calculate the shock induced global
accretion solutions and investigate the effect of dissipation parameters, such
as ${\dot m}$, $\alpha_B$ and $\beta$, in deciding the formation of shock waves.
The results are summarized below.

We find that accreting matter continues to harbor standing shock waves for
$a_k \leq 0.8$ (see Fig. 1-5). It may be noted that we restrict the upper
limit of $a_k$ below its maximum allowed value ($i. e., a_k \rightarrow 1$),
because the adopted potential satisfactorily mimics the space-time geometry
around the rotating black hole for spin parameter $a_k \lesssim 0.8$ 
\citep{Chakrabarti-Mondal06}. Furthermore, we have realized that standing shocks
in magnetized accretion flow are 
quite common and they exist for a wide range of flow parameters (see Fig. 2-5). 

Next, we quantify the range of dissipation parameters that admit the
formation of standing shocks in magnetized accretion flow around
rotating black holes. We find that flow can sustain shock waves even
when the level of dissipation is very high. More importantly, we observe
that radiative cooling acts oppositely in contrast with viscous dissipation
in deciding the shock parameter space (see Fig. 6). However, the effect 
of cooling can not be mitigated completely by viscous heating as their
dependencies on the flow variables are different.
Further, we find that the possibility of shock formation
always decreases with the increase of dissipation strength.
Subsequently, we calculate the critical accretion rate ($\dot{m}^{\rm cri}$)
for standing shock. When accretion rate exceeds the critical limit,
standing shock conditions are not satisfied and consequently, standing shock
disappears. We find that $\dot{m}^{\rm cri}$ strongly depends on viscosity
($\alpha_B$), magnetic fields ($\beta$) and spin of the black hole ($a_k$),
respectively (see Fig. 6-9). 
What is more is that standing shock exists in a magnetically dominated
accretion flow when the accretion rate lies in general in the sub-Eddington
domain (${\dot m} < 1$) whereas for gas pressure dominated flow, shock
forms even for super-Eddington accretion rate (${\dot m} > 1$) (see Fig. 7-9).

Further, we obtain the standing shock solution for magnetized
accretion flow, where the shock is considered to be dissipative by nature. The
available energy dissipated at shock ($\Delta \mathcal{E}$) is usually escaped
through the disc surface that is being utilized to power the
jets/outflows \citep{Le-Becker04,Le-Becker05,Das-etal09}. Towards this,
we compute the maximum energy dissipated at shock 
($\Delta{\cal E}^{\rm max}$) and find that $\Delta{\cal E}^{\rm max}$ increases 
with $a_k$ although its dependence on $\beta_{\rm in}$ is very much
conspicuous.

Finally, we would like to mention that the present formalism is developed by
adopting a simplified pseudo potential 
to delineate the gravitational effect around a rotating black
hole. Incidentally, while studying the non-linear shock solutions, this
approach allow us to avoid the mathematical complexity of general
theory of relativity and at the same time it retains the salient features
of space-time geometry around rotating black holes 
\citep{Chakrabarti-Mondal06}. In this regard, although 
the present formalism introduces a bit of imperfections, however,
we believe that the basic findings of this work will qualitatively remain
unaltered due to this approximation.

\section*{Acknowledgments}
Authors thank the anonymous referee for useful comments and constructive suggestions.



\begin{thebibliography}{reference}

\bibitem[\protect\citeauthoryear{Aitken et al.}{1993}]{Aitken-etal93} 
Aitken D.~K., Wright C.~M., Smith C.~H., Roche P.~F., 1993, MNRAS, 262, 456

\bibitem[\protect\citeauthoryear{Akizuki \& Fukue}{2006}]{Akizuki-Fukue06} 
Akizuki C., Fukue J., 2006, PASJ, 58, 469 

\bibitem[\protect\citeauthoryear{Aktar, Das, \& Nandi}{2015}]{Aktar-etal15} 
Aktar R., Das S., Nandi A., 2015, MNRAS, 453, 3414 

\bibitem[\protect\citeauthoryear{Aktar et al.}{2017}]{Aktar-etal17} 
Aktar R., Das S., Nandi A., Sreehari H., 2017, MNRAS, 471, 4806 

\bibitem[\protect\citeauthoryear{Balbus \& Hawley}{1991}]{Balbus-Hawley91} 
Balbus S.~A., Hawley J.~F., 1991, ApJ, 376, 214 

\bibitem[\protect\citeauthoryear{Balbus \& Hawley}{1998}]{Balbus-Hawley98} 
Balbus S.~A., Hawley J.~F., 1998, RvMP, 70, 1 

\bibitem[\protect\citeauthoryear{Blandford \& Payne}{1982}]{Blandford-Payne82} 
Blandford R.~D., Payne D.~G., 1982, MNRAS, 199, 883 

\bibitem[\protect\citeauthoryear{Blandford \& Znajek}{1977}]{Blandford-Znajek77} 
Blandford R.~D., Znajek R.~L., 1977, MNRAS, 179, 433 

\bibitem[\protect\citeauthoryear{Becker \& Kazanas}{2001}]{Becker-Kazanas01} 
Becker P.~A., Kazanas D., 2001, ApJ, 546, 429 

\bibitem[\protect\citeauthoryear{Becker, Das, \& Le}{2008}]{Becker-etal2008} 
Becker P.~A., Das S., Le T., 2008, ApJ, 677, L93 

\bibitem[\protect\citeauthoryear{Chakrabarti}{1989}]{Chakrabarti89} 
Chakrabarti S.~K., 1989, ApJ, 347, 365 

\bibitem[\protect\citeauthoryear{Chakrabarti \&
Molteni}{1993}]{ChakrabartiMolteni1993} 
Chakrabarti S.~K., Molteni D., 1993, ApJ, 417, 671

\bibitem[\protect\citeauthoryear{Chakrabarti \&
Titarchuk}{1995}]{Chakrabarti-Titarchuk95} 
Chakrabarti S., Titarchuk L.~G., 1995, ApJ, 455, 623 

\bibitem[\protect\citeauthoryear{Chakrabarti}{1996a}]{Chakrabarti96a} 
Chakrabarti S.~K., 1996a, ApJ, 464, 664 

\bibitem[\protect\citeauthoryear{Chakrabarti}{1996b}]{Chakrabarti96b} 
Chakrabarti S.~K., 1996b, MNRAS, 283, 325

\bibitem[\protect\citeauthoryear{Chakrabarti \& Das}{2004}]{Chakrabarti-Das04} 
Chakrabarti S.~K., Das S., 2004, MNRAS, 349, 649

\bibitem[\protect\citeauthoryear{Chakrabarti \& Mondal}{2006}]{Chakrabarti-Mondal06} 
Chakrabarti S.~K., Mondal S., 2006, MNRAS, 369, 976 
 
\bibitem[\protect\citeauthoryear{Chattopadhyay \&
Chakrabarti}{2000}]{Chattopadhyay-Chakrabarti00} 
Chattopadhyay I., Chakrabarti S.~K., 2000, IJMPD, 9, 717 

\bibitem[\protect\citeauthoryear{Chattopadhyay \&
Chakrabarti}{2002}]{Chattopadhyay-Chakrabarti02} 
Chattopadhyay I., Chakrabarti S.~K., 2002, MNRAS, 333, 454 

\bibitem[\protect\citeauthoryear{Chuss et al.}{2003}]{Chuss-etal03} 
Chuss D.~T., Davidson J.~A., Dotson J.~L., Dowell C.~D., Hildebrand R.~H., Novak G.,
Vaillancourt J.~E., 2003, ApJ, 599, 1116 

\bibitem[\protect\citeauthoryear{Das, Chattopadhyay, \&
Chakrabarti}{2001a}]{Das-etal01a} 
Das S., Chattopadhyay I., Chakrabarti S.~K., 2001a, ApJ, 557, 983 

\bibitem[\protect\citeauthoryear{Das et al.}{2001b}]{Das-etal01b} 
Das S., Chattopadhyay I., Nandi A., Chakrabarti S.~K., 2001b, A\&A, 379, 683

\bibitem[\protect\citeauthoryear{Das}{2007}]{Das07} 
Das S., 2007, MNRAS, 376, 1659

\bibitem[\protect\citeauthoryear{Das \& Chakrabarti}{2008}]{Das-Chakrabarti08} 
Das S., Chakrabarti S.~K., 2008, MNRAS, 389, 371 

\bibitem[\protect\citeauthoryear{Das, Becker, \& Le}{2009}]{Das-etal09} 
Das S., Becker P.~A., Le T., 2009, ApJ, 702, 649 

\bibitem[\protect\citeauthoryear{Das, Chakrabarti, \& Mondal}{2010}]{Das-etal10} 
Das S., Chakrabarti S.~K., Mondal S., 2010, MNRAS, 401, 2053 

\bibitem[\protect\citeauthoryear{Das \& Aktar}{2015}]{Das-Aktar2015} 
Das S., Aktar R., 2015, ASInC, 12, 

\bibitem[\protect\citeauthoryear{Fukue}{1987}]{Fukue87} 
Fukue J., 1987, PASJ, 39, 309

\bibitem[\protect\citeauthoryear{Fukue}{1990}]{Fukue90} 
Fukue J., 1990, PASJ, 42, 793

\bibitem[\protect\citeauthoryear{Fukumura \& Tsuruta}{2004}]{Fukumura-Tsuruta04} 
Fukumura K., Tsuruta S., 2004, ApJ, 611, 964 

\bibitem[\protect\citeauthoryear{Fukumura, Takahashi, \&
Tsuruta}{2007}]{Fukumura-etal07} 
Fukumura K., Takahashi M., Tsuruta S., 2007, ApJ, 657, 415 

\bibitem[\protect\citeauthoryear{Fukumura et al.}{2016}]{Fukumura-etal16} 
Fukumura K., Hendry D., Clark P., Tombesi F., Takahashi M., 2016, ApJ, 827, 31 

\bibitem[\protect\citeauthoryear{Gu \& Lu}{2001}]{Gu-Lu01} 
Gu W.-M., Lu J.-F., 2001, ChPhL, 18, 148 

\bibitem[\protect\citeauthoryear{Gu \& Lu}{2004}]{Gu-Lu04} 
Gu W.-M., Lu J.-F., 2004, ChPhL, 21, 2551 

\bibitem[\protect\citeauthoryear{Hirose, Krolik, \& Stone}{2006}]{Hirose-etal06} 
Hirose S., Krolik J.~H., Stone J.~M., 2006, ApJ, 640, 901 

\bibitem[\protect\citeauthoryear{Johansen \& Levin}{2008}]{Johansen-Levin08} 
Johansen A., Levin Y., 2008, A\&A, 490, 501 

\bibitem[\protect\citeauthoryear{Khesali \& Faghei}{2008}]{Khesali-Faghei08} 
Khesali A., Faghei K., 2008, MNRAS, 389, 1218

\bibitem[\protect\citeauthoryear{Khesali \& Faghei}{2009}]{Khesali-Faghei09} 
Khesali A., Faghei K., 2009, MNRAS, 398, 1361 

\bibitem[\protect\citeauthoryear{Komissarov \& McKinney}{2007}]{Komissarov-McKinney07} 
Komissarov S.~S., McKinney J.~C., 2007, MNRAS, 377, L49 

\bibitem[\protect\citeauthoryear{Kumar \& Chattopadhyay}{2013}]{kc13} 
Kumar R., Chattopadhyay I., 2013, MNRAS, 430, 386 

\bibitem[\protect\citeauthoryear{Le \& Becker}{2004}]{Le-Becker04} 
Le T., Becker P.~A., 2004, ApJ, 617, L25 

\bibitem[\protect\citeauthoryear{Le \& Becker}{2005}]{Le-Becker05} 
Le T., Becker P.~A., 2005, ApJ, 632, 476 

\bibitem[\protect\citeauthoryear{Lu, Gu, \& Yuan}{1999}]{Lu-etal99} 
Lu J.-F., Gu W.-M., Yuan F., 1999, ApJ, 523, 340 

\bibitem[\protect\citeauthoryear{Machida, Nakamura, \&
Matsumoto}{2006}]{Machida-etal06} 
Machida M., Nakamura K.~E., Matsumoto R., 2006, PASJ, 58, 193 

\bibitem[\protect\citeauthoryear{Matsumoto et al.}{1984}]{Matsumoto-etal84} 
Matsumoto R., Kato S., Fukue J., Okazaki A.~T., 1984, PASJ, 36, 71 

\bibitem[\protect\citeauthoryear{Molteni, Lanzafame, \&
Chakrabarti}{1994}]{Moltenietal1994} 
Molteni D., Lanzafame G., Chakrabarti S.~K., 1994, ApJ, 425, 161 

\bibitem[\protect\citeauthoryear{Mondal \& Chakrabarti}{2006}]{Mondal-Chakrabarti06} 
Mondal S., Chakrabarti S.~K., 2006, MNRAS, 371, 1418 

\bibitem[\protect\citeauthoryear{Mosallanezhad, Abbassi, \&
Beiranvand}{2014}]{mosall14} 
Mosallanezhad A., Abbassi S., Beiranvand N., 2014, MNRAS, 437, 3112 

\bibitem[\protect\citeauthoryear{Mosallanezhad, Bu, \& Yuan}{2016}]{mosall16} 
Mosallanezhad A., Bu D., Yuan F., 2016, MNRAS, 456, 2877 

\bibitem[\protect\citeauthoryear{Narayan \& Yi}{1995}]{Narayan-Yi95} 
Narayan R., Yi I., 1995, ApJ, 452, 710

\bibitem[\protect\citeauthoryear{Nishikawa et al.}{2005}]{Nishikawa-etal05} 
Nishikawa K.-I., Richardson G., Koide S., Shibata K., Kudoh T., Hardee P., Fishman
G.~J., 2005, ApJ, 625, 60 

\bibitem[\protect\citeauthoryear{Novak et al.}{2003}]{Novak-etal03} 
Novak G., et al., 2003, ApJ, 583, L83 

\bibitem[\protect\citeauthoryear{Okuda}{2014}]{Okuda14} 
Okuda T., 2014, MNRAS, 441, 2354 

\bibitem[\protect\citeauthoryear{Okuda \& Das}{2015}]{Okuda-Das15} 
Okuda T., Das S., 2015, MNRAS, 453, 147 

\bibitem[\protect\citeauthoryear{Oda et al.}{2007}]{Oda-etal07} 
Oda H., Machida M., Nakamura K.~E., Matsumoto R., 2007, PASJ, 59, 457

\bibitem[\protect\citeauthoryear{Oda et al.}{2010}]{Oda-etal10} 
Oda H., Machida M., Nakamura K.~E., Matsumoto R., 2010, ApJ, 712, 639 

\bibitem[\protect\citeauthoryear{Oda et al.}{2012}]{Oda-etal12} 
Oda H., Machida M., Nakamura K.~E., Matsumoto R., Narayan R., 2012, PASJ, 64, 15 

\bibitem[\protect\citeauthoryear{Ryu, Chakrabarti, \& Molteni}{1997}]{Ryuetal1997} 
Ryu D., Chakrabarti S.~K., Molteni D., 1997, ApJ, 474, 378

\bibitem[\protect\citeauthoryear{Samadi, Abbassi, \& Khajavi}{2014}]{Samadi-etal14} 
Samadi M., Abbassi S., Khajavi M., 2014, MNRAS, 437, 3124

\bibitem[\protect\citeauthoryear{Sarkar \& Das}{2013}]{sarkar2013} 
Sarkar B., Das S., 2013, ASInC, 8, 143 

\bibitem[\protect\citeauthoryear{Sarkar \& Das}{2015}]{sarkar2015} 
Sarkar B., Das S., 2015, ASInC, 12, 

\bibitem[\protect\citeauthoryear{Sarkar \& Das}{2016}]{sarkar2016} 
Sarkar B., Das S., 2016, MNRAS, 461, 190 

\bibitem[\protect\citeauthoryear{Sarkar \& Das}{2018}]{sarkar2017b} 
Sarkar B., Das S., 2018, JApA, 39, 3 

\bibitem[\protect\citeauthoryear{Sarkar, Das, \& Mandal}{2018}]{sarkar2018} 
Sarkar B., Das S., Mandal S., 2018, MNRAS, 473, 2415 

\bibitem[\protect\citeauthoryear{Shakura \& Sunyaev}{1973}]{Shakura-Sunyaev73} 
Shakura N.~I., Sunyaev R.~A., 1973, A\&A, 24, 337

\bibitem[\protect\citeauthoryear{Shapiro \&
Teukolsky}{1983}]{Shapiro-Teukolsky83}
Shapiro S. L., Teukolsky S. A., 1983, Black Holes, White Dwarfs and 
Neutron Stars: The Physics of Compact Objects. Wiley, New York

\bibitem[\protect\citeauthoryear{Singh \& Chakrabarti}{2011}]{Singh-Chakrabarti11} 
Singh C.~B., Chakrabarti S.~K., 2011, MNRAS, 410, 2414 

\bibitem[\protect\citeauthoryear{Sukov{\'a} \& Janiuk}{2015}]{Sukova-Janiuk15} 
Sukov{\'a} P., Janiuk A., 2015, MNRAS, 447, 1565 

\bibitem[\protect\citeauthoryear{Sukov{\'a}, Charzy{\'n}ski, \&
Janiuk}{2017}]{Sukova-17} 
Sukov{\'a} P., Charzy{\'n}ski S., Janiuk A., 2017, MNRAS, 472, 4327 

\bibitem[\protect\citeauthoryear{Takahashi et al.}{2006}]{Takahashi-etal06} 
Takahashi M., Goto J., Fukumura K., Rilett D., Tsuruta S., 2006, ApJ, 645, 1408 

\bibitem[\protect\citeauthoryear{Wright, Aitken, \& Smith}{1993}]{Wright-etal93} 
Wright C.~M., Aitken D.~K., Smith C.~H., 1993, PASAu, 10, 247

\end{thebibliography}
\end{document}